\documentclass[intlimits,twoside,a4paper]{article}

\usepackage[cp1251]{inputenc}
\usepackage{multirow}
\usepackage{xcolor}
\usepackage[eqsecnum]{cmpj3}

\usepackage{bm}

\issue{2023}{26}{2}{23707}
\doinumber{10.5488/CMP.26.23707}

\title[Topological phase transition in SrSi$_{2}$]%
{Pressure driven Weyl-topological insulator phase transition in Weyl semimetal SrSi$_{2}$%
}
\author[Aditya Shende, Shivendra Kumar Gupta, Ashish Kore, Poorva Singh]{Aditya Shende\orcid{0000-0003-2231-5946}, Shivendra Kumar Gupta\orcid{0000-0002-1631-5846}, Ashish Kore\orcid{0000-0002-3649-9718}, Poorva Singh\orcid{0000-0002-8133-7945}\thanks{Corresponding author: \email{poorvasingh@phy.vnit.ac.in}}}

\address{Department of Physics, Visvesvaraya National Institute of Technology,  S Ambazari
Road, \\ Nagpur, 440010, Maharashtra, India
}

\date{Received January 1, 2023, in final form March 24, 2023}

\begin{document}
	
	\maketitle

\begin{abstract}
Using DFT-based first-principles calculations, we demonstrate the tuning of the electronic structure of Weyl semimetal SrSi$_{2}$ via  external  uniaxial  strain. The  uniaxial  strain  facilitates the  opening  of bandgap along $\Gamma$-X direction and subsequent band inversion between Si $p$ and Sr $d$ orbitals. Z$_{2}$ invariants and surface states reveal conclusively that  SrSi$_{2}$ under uniaxial strain is a strong topological insulator.   Hence, uniaxial strain drives the semimetallic SrSi$_{2}$ into fully gapped topological insulating state depicting a semimetal to topological insulator phase transition. Our results highlight the suitability of uniaxial strain to gain control over the topological phase transitions and topological states in SrSi$_{2}$.
%
%
\printkeywords
%
\end{abstract}

\section{Introduction}

The discovery of topological insulators (TI) \cite{ref1,ref2,ref3} and topological semimetals (TSM) \cite{ref4,ref5,ref6,ref7} has revolutionized the field of condensed matter  physics. Topological insulators exhibit exotic metallic surface states in conjunction with the gapped bulk state while topological semimetals are characterized by valence and conduction  band overlaps located precisely at the Fermi level. The topological state was initially anticipated in two-dimensional HgTe/CdTe quantum wells \cite{ref8} and afterwards affirmed by ARPES experiments \cite{ref9}.  Later, two-dimensional conducting surface states were realized in the bulk band gap of three-dimensional materials Bi$_{1-x}$Sb$_{x}$, Bi$_{2}$Se$_{3}$, Bi$_{2}$Te$_{3}$ and Sb$_{2}$Te$_{3}$ \cite{ref10,ref11}. Predictions from first principles calculations have proposed a huge number of topologically non-trivial materials extending from oxides \cite{ref12} to the Heusler family of compounds \cite{ref13,ref14,ref15} and thallium-based ternary III-V-VI2 chalcogenides \cite{ref16}.
TSMs have roused up  immense excitement due to their proposed application in synthesis catalysis, quantum computation, and spintronics. In TSMs, guided by lattice symmetry and topology, different band overlap  scenarios may emerge, e.g., a Dirac point or Weyl point or nodal line/nodal ring may be formed on the Fermi surface. The corresponding material classes are known as Dirac or Weyl semimetal (DSM/WSM) or nodal line semimetal, respectively \cite{ref7}. Due to the nontrivial topology of the bulk and surface electronic states, TSMs are likely to show exciting quantum transport phenomena, such as unusual magneto-resistance and chiral anomaly. Recent material realizations of TSMs include  Dirac semimetal Na$_{3}$Bi \cite{ref5},  Cd$_{3}$As$_{2}$ \cite{ref17}; Weyl semimetal TaAs \cite{ref18,ref19,ref20,ref21,ref22}, SrSi$_{2}$ \cite{ref23,ref24}; magnetic pyrochlores A$_{2}$Ir$_{2}$O$_{7}$ \cite{ref6}; ferromagnetic half-metal HgCr$_{2}$Se$_{4}$ \cite{ref25,ref26}; triple-point semimetals MoP \cite{ref27} and WC \cite{ref28}, and the Dirac nodal line semimetal ZrSiS \cite{ref29}. 

Topological order may be tuned via changing the chemical composition and thereby modifying both 
the strength of spin orbit coupling (SOC) and the lattice parameters. Another strategy is to apply external strain \cite{ref30,ref31} which has been theoretically shown for a number of narrow band gap cubic semiconductors such as grey tin ($\alpha$-Sn) \cite{PhysRevB.91.035311} and HgTe \cite{PhysRevB.91.035311, PhysRevLett.106.126803}, InSb \cite{PhysRevB.85.195114}, KNa$_{2}$Bi \cite{Sklyadneva2016}, TaAs \cite{PhysRevLett.117.146402}, Co$_{3}$Sn$_{2}$S$_{2}$\cite{ref35}. Hence, compressive/tensile  strain is utilised to close/reopen the bandgap in some topological materials, which may be accompanied by strain-induced topological phase transitions \cite{ref30,ref32,ref33,ref34}. Recent studies indicate that SrSi$_{2}$ is a robust double-Weyl semimetal due to the absence of inversion symmetry. SrSi$_{2}$ is WSM in its ground state even without the requirement of spin-orbit coupling. SrSi$_{2}$ is also predicted to undergoes topological phase transition from Weyl semimetal to trivial insulator by increasing its lattice constant uniformly in all three directions \cite{ref23}. Hence, triaxial strain is unsuitable to retain the topological behaviour in SrSi$_{2}$. In comparison to triaxial strain,  the application of uniaxial strain may offer an additional advantage in terms of introducing asymmetry into the lattice. It was earlier  shown that uniaxial strain \cite{ref30, Sklyadneva2016,PhysRevB.102.035132} is an effective tool to realize a topological insulating state by introducing a bulk band gap. Hence,  using the uniaxial strain, it is possible to gain better control of topological properties and reveal topological phase transitions of SrSi$_{2}$.
\begin{figure}
	\centering
	\includegraphics[angle=-0.0,origin=c,height=6.6cm,width=10.5cm,scale=1.5]{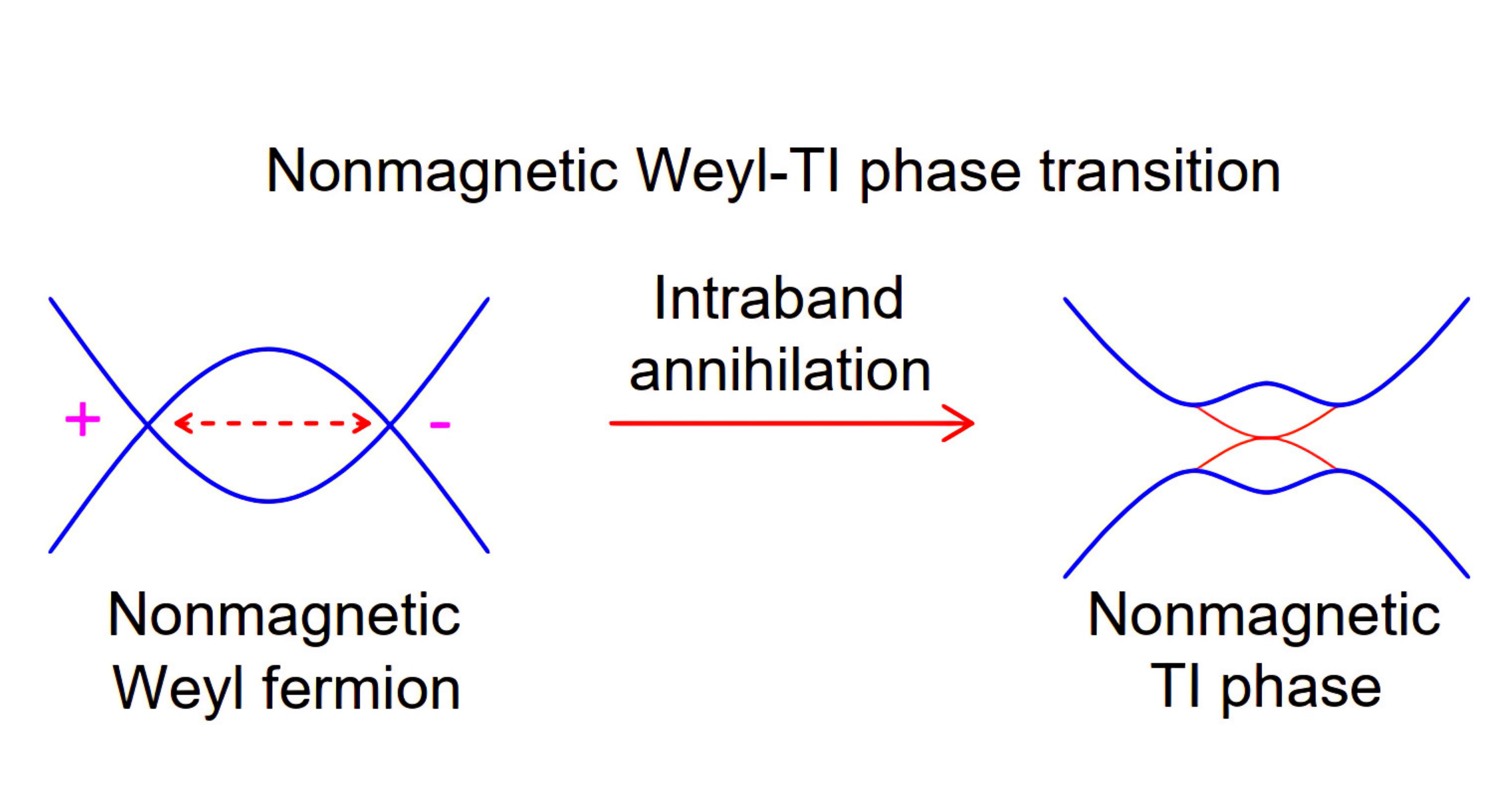}
	\caption{(Colour online) Weyl-TI TPT in a non-magnetic system with intraband annihilation of chiral Weyl node \cite{ref35}}
	\label{fig1}
\end{figure}

In this work, we employ first principle  calculations to   study the effect of applying uniaxial compressive strain on SrSi$_{2}$. We discover a topological semimetal to topological insulator quantum phase transition and obtain a fully gapped quantum spin Hall state (QSH) in SrSi$_{2}$ at 5\%  strain. Here, we consider the new Weyl-TI phase transition which includes two separate topological non-trivial phases as well as the metal-insulator transition. The chiral Weyl nodes annihilate in pairs during phase transition if time-reversal symmetry is preserved \cite{ref35} (figure~\ref{fig1}).  We predict a new \textit{d-p} band inversion in SrSi$_{2}$ using band structure calculations. We confirm the stability of a strained structure using phonon band structure and also compute the Z$_{2}$ invariants and surface band structure to confirm its topological insulating behaviour.
The organization of this paper is as follows. In section~\ref{sec2} we provide the crystal structure of bulk SrSi$_{2}$ and details of computational methodology. Section~\ref{sec3}  is split into four parts. In section~\ref{sec3.1}, electronic band structures  of bulk SrSi$_{2}$ are studied. In section~\ref{sec3.2}, strain-influenced structures of SrSi$_{2}$ and their electronic band structures  are studied. In section~\ref{sec3.3}, we discuss the band inversion in the strain-influenced structures and confirm their topological insulating behaviour by computing Z$_{2}$ invariants. Later in section~\ref{sec3.4}, we discuss the surface states of the strain-influenced structures. In section~\ref{sec4},  we discuss the primary conclusions drawn from the present work.

\section{Crystal structure and computational methodology}\label{sec2}
\begin{figure}[!t]
 \centering
{\includegraphics[width=0.45\textwidth]{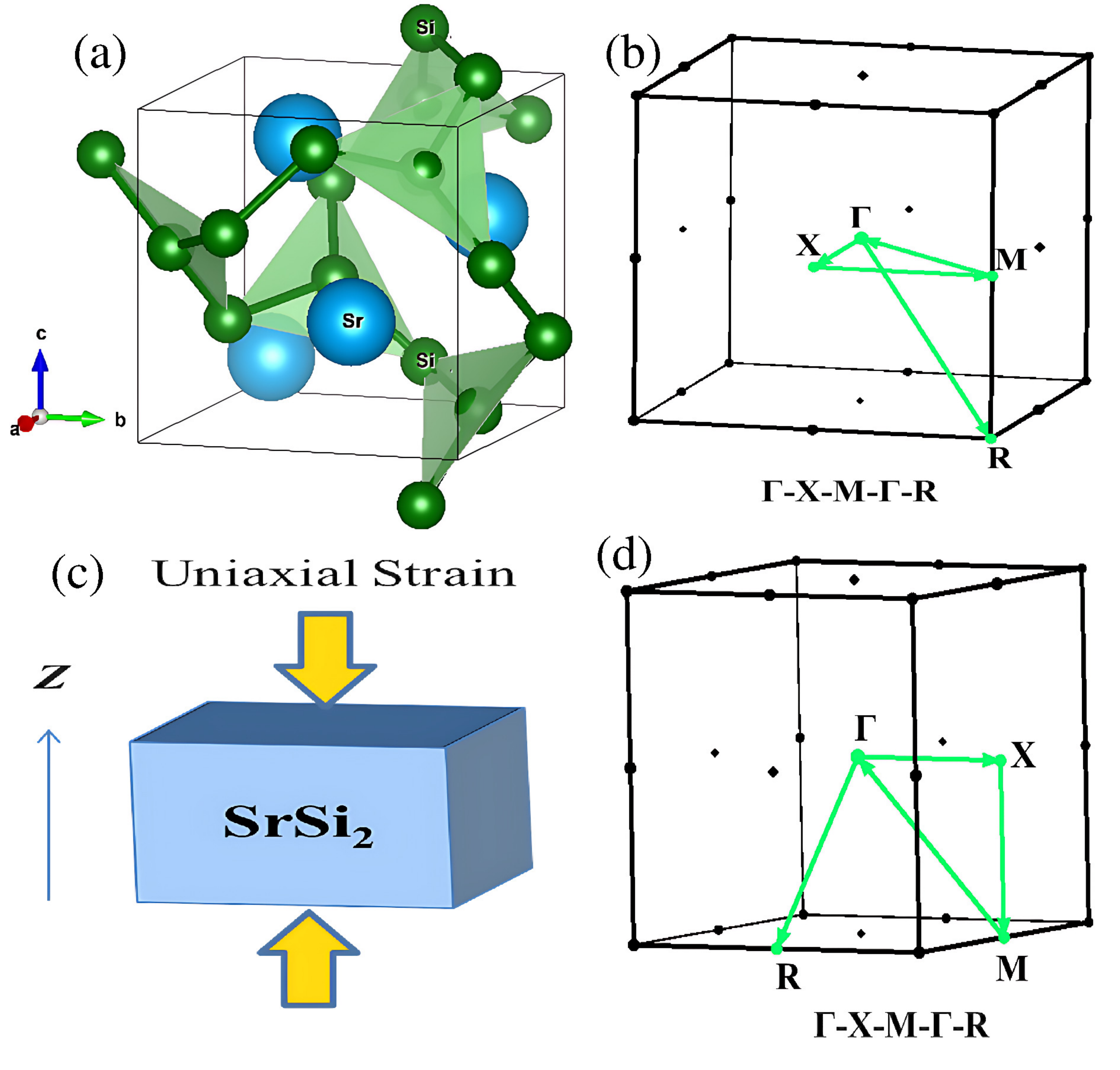}}
  \caption{(Colour online) (a) Primitive unit cell of bulk SrSi$_{2}$ (cubic lattice)  in the ground state, (b) the corresponding Brillioun zone with high symmetry points, (c) SrSi$_{2}$ under the uniaxial strain  (tetragonal lattice) along z-direction and (d) the corresponding Brillioun zone with high symmetry points.}
  \label{fig2}
\end{figure}

SrSi$_{2}$ belongs to the cubic crystal system with the space group P4$_{3}$32 (No.~212). Figure~\ref{fig2}(a) indicates the primitive cell of SrSi$_{2}$ in the ground state which involves four atoms of Sr and eight atoms of Si, each  Si atom is surrounded by three atoms of Si  and seven atoms of Sr, whereas each atom of Sr  is in the vicinity of fourteen atoms of Si as well as six atoms of Sr. The positioning of the atoms in the ground state of SrSi$_{2}$ are at Sr: 4a (0.125, 0.125, 0.125) and Si: 8c (0.43, 0.43, 0.43). Figure~\ref{fig2}(c) indicates the schematic diagram of SrSi$_{2}$ under uniaxial strain. Figure~\ref{fig2}(b) and (c) denotes the Brillioun zone for ground and strained lattices, respectively.

For bulk and strained SrSi$_{2}$, the density functional theory (DFT) based calculations were performed using Quantum Espresso code \cite{ref36}, with the standard frozen-core projector augmented-wave (PAW) method. The exchange correlation of generalized gradient approximation in the  Perdew-Burke-Ernzerhof (GGA-PBE) format was employed. We used kinetic energy cut-off for wave function as 45 Ry and $12 \times 12 \times 12$ 
\textit{k}-point grid to sample the Brillouin zone for optimizing the crystal structure and self-consistent calculations. For the calculation of topological invariants and the surface states, we used Wannier wave functions \cite{ref37,ref38,ref39} as implemented in the Wannier90 package \cite{ref40}. Phonon calculation is done using the  phonopy code and $2 \times 2 \times 2$ supercell to obtain the phonon dispersion by means of density functional perturbation theory method \cite{ref41}.
\begin{figure}[!t]
	\centering
	\includegraphics[width=0.8\textwidth]{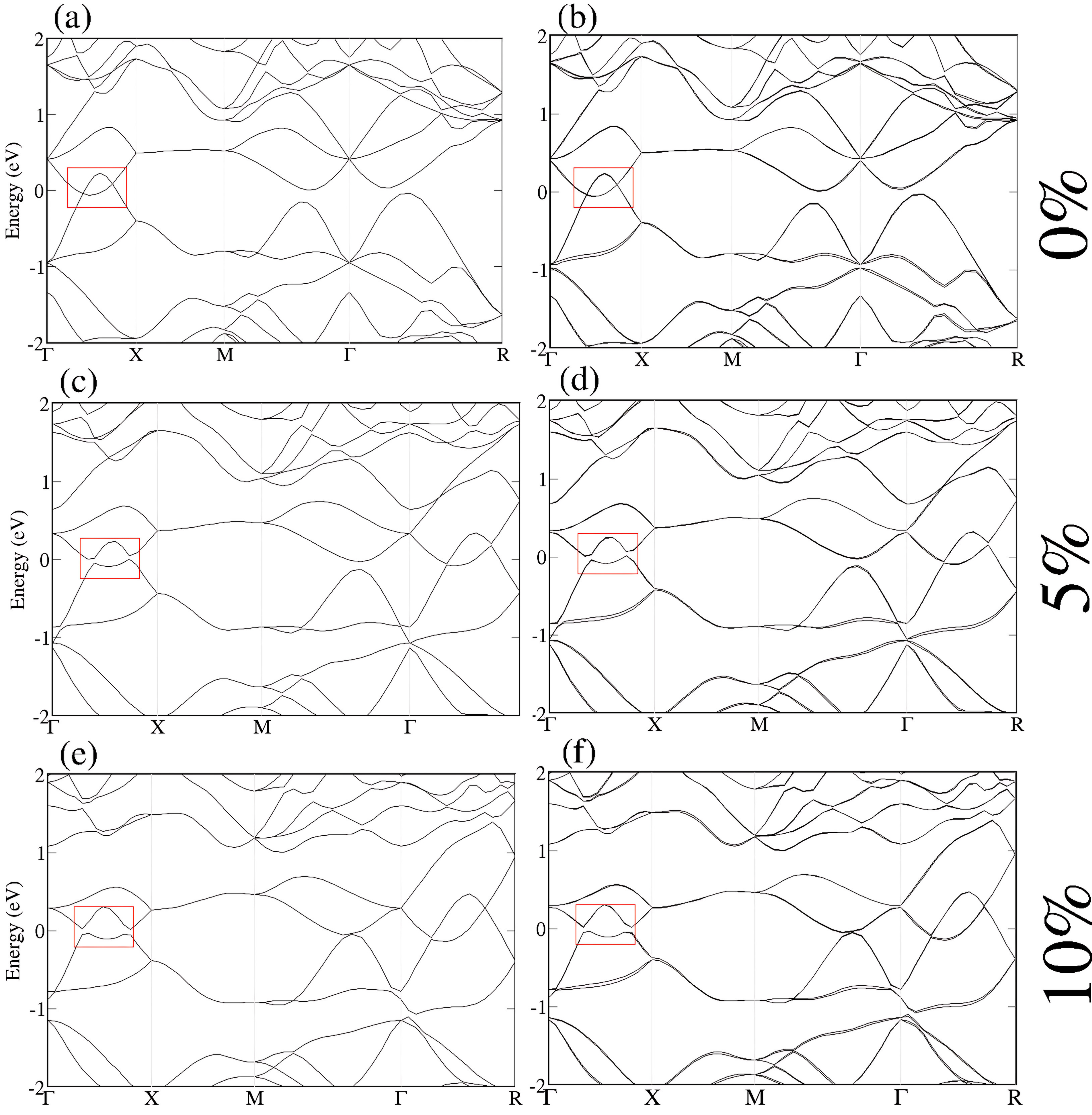}
	\caption{(Colour online) The computed electronic band structure of bulk SrSi$_2$ across high symmetry points obtained from a) GGA and b) GGA+SOC for the ground state, c) GGA and d) GGA+SOC for 5\% strained structure of SrSi$_2$, e) GGA and f) GGA+SOC for 10\% strained structure of SrSi$_2$.}
	\label{fig3}
	\centering
\end{figure}

\begin{figure}[!t]
	\centering
	{\includegraphics[width=0.9\textwidth]{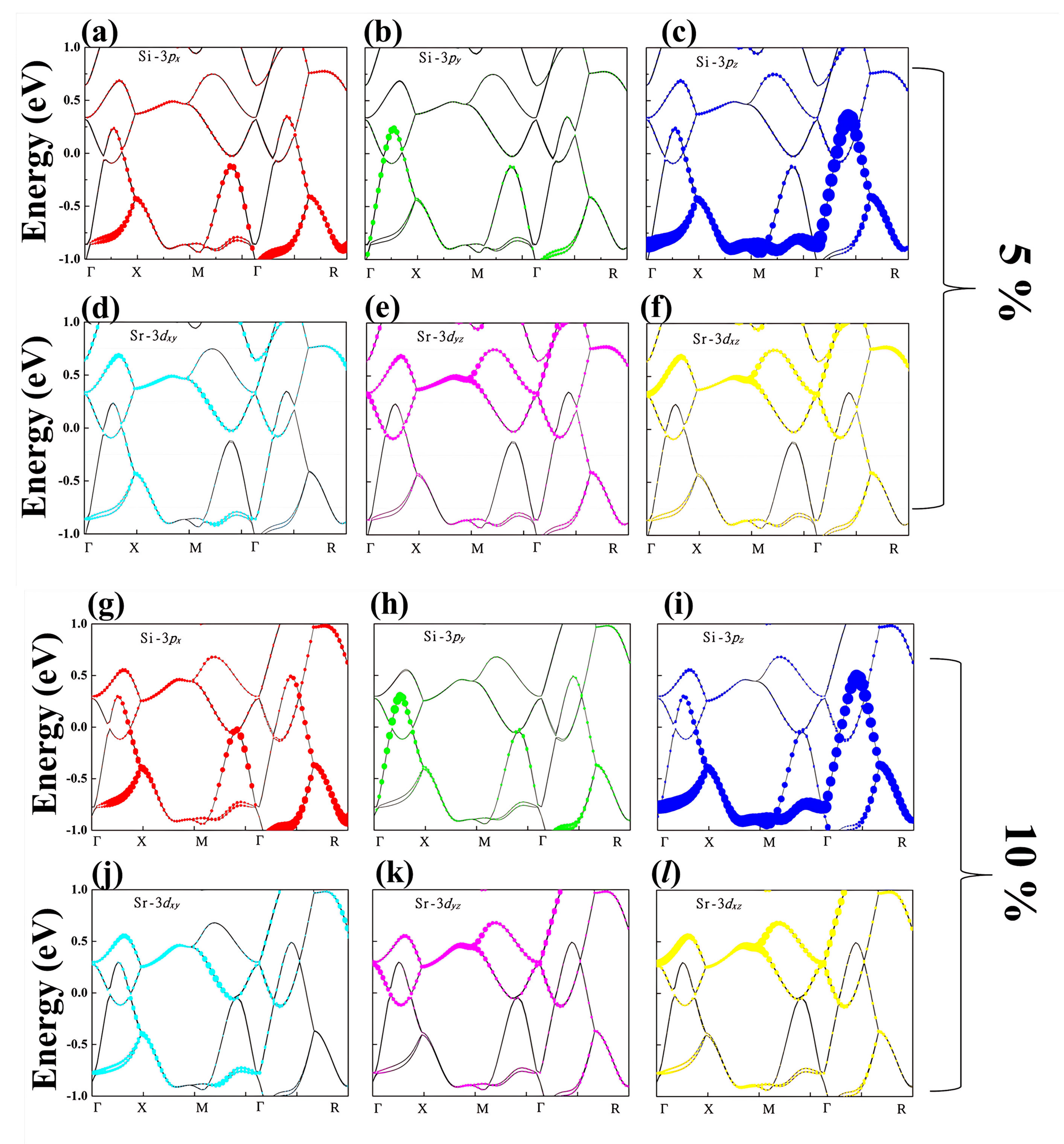}\label{first}}
	\caption{(Colour online) (a)--(f) Orbital projected band structures for 5\% strain and (g)--(l) for 10\% strain on SrSi$_2$. The radii of the circles depict the weights of the corresponding orbital and the red, green and blue colours correspond to $p_{x}$, $p_{y}$ and $p_{z}$ orbitals of Si, while the cyan, magenta and yellow colours correspond to d$_{xy}$, $d_{yz}$ and $d_{xz}$ orbitals of Sr.}
	\label{main_label}
\end{figure}

\begin{figure}[!t]%
	\centering
	{\includegraphics[width=0.6\textwidth]{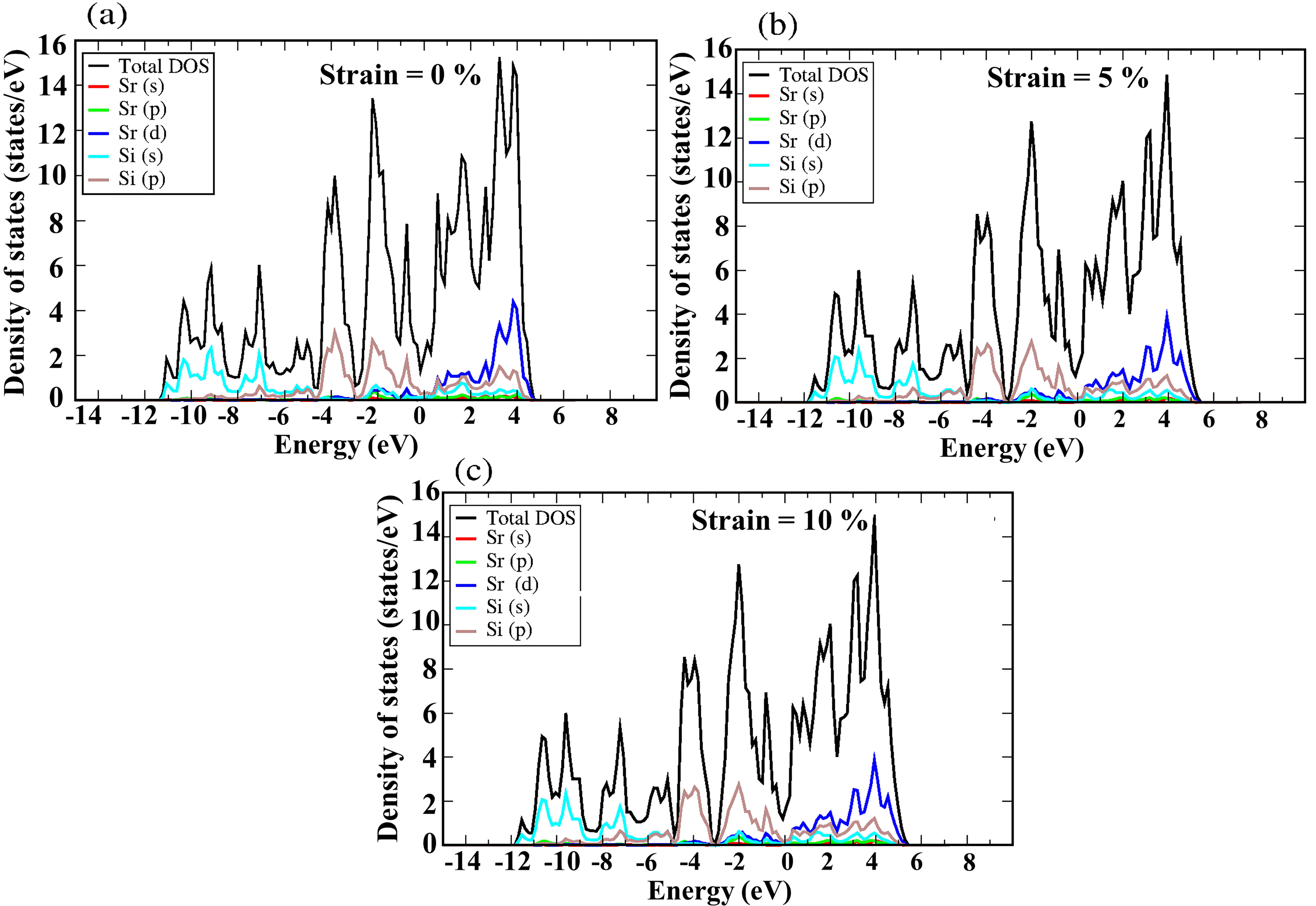}\label{fig_5}}
	\caption{(Colour online) The total and partial densities of states at (a) 0\% (b) 5 \% and (c) 10\% strain on SrSi$_2$.
	}
	\label{some-label}%
\end{figure}
\begin{figure}[!h]
	\centering
	\includegraphics[width=0.6\textwidth]{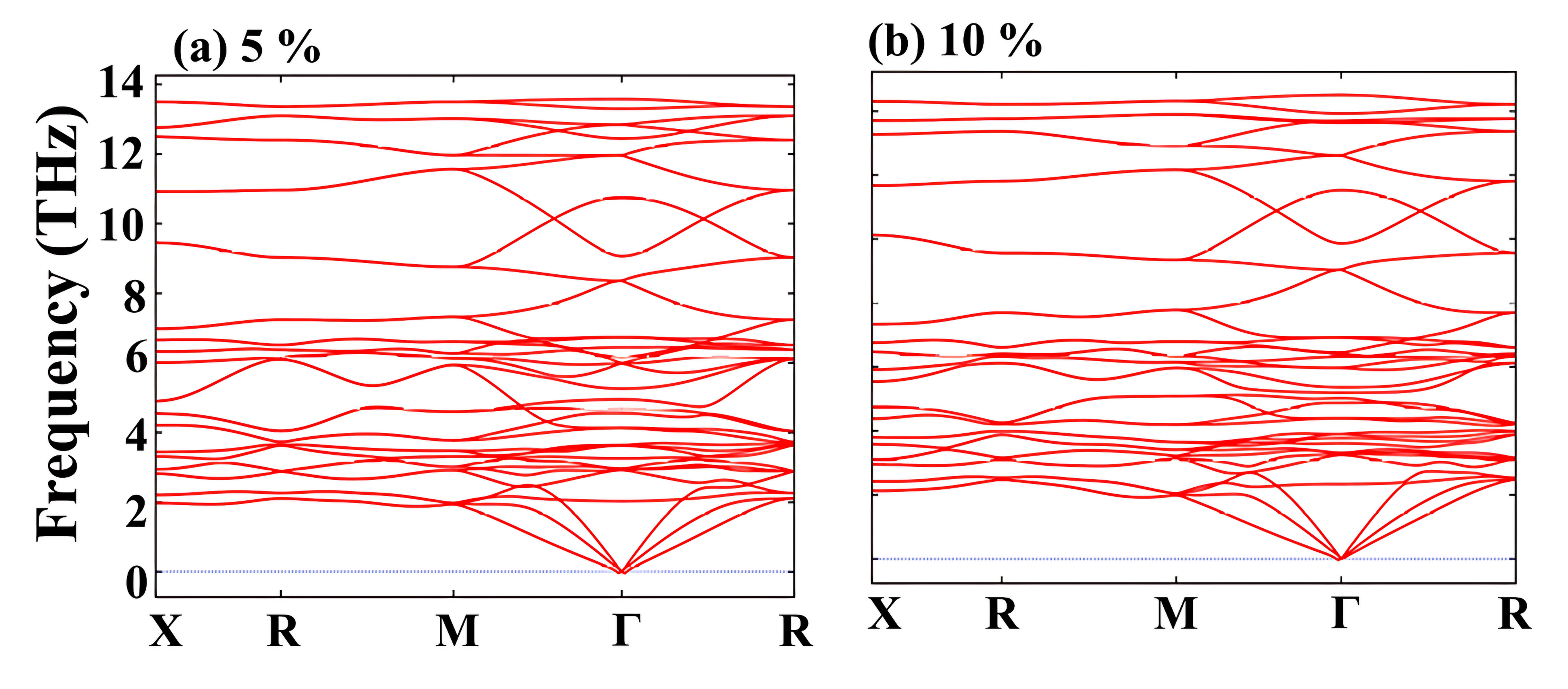}
	\caption{(Colour online) Phonon dispersion for (a) 5\% and (b) 10\% strained structures of  SrSi$_{2}$.}
	\label{fig6}
\end{figure} 
\section{Results and discussion}\label{sec3}
\subsection{Bulk SrSi$_{2}$}\label{sec3.1}
The entire crystal volume was optimized in order to attain ground state  lattice parameters for bulk SrSi$_{2}$. The energy-volume ($E-V$) data was then fitted in the Birch-Murnaghan equation of state. The optimized lattice parameter was found to be $a = 6.563$~{\AA},  in good agreement with previous experimental~\cite{ref42,ref42ol} and theoretical \cite{ref23} results. For the ground state bulk SrSi$_{2}$, the electronic band structure  from GGA and GGA+SOC (spin-orbit coupling) along high symmetry path was calculated as  shown in figure~\ref{fig3}(a) and (b). In consonance with previous reports, we notice that the valance band (VB) and conduction band (CB) touch each other at two distinct k points along $\Gamma$-X direction,  clearly revealing the semimetallic nature of SrSi$_{2}$. Our computed energy band structures are in good agreement with the previous results \cite{ref23,ref24}. Due to the absence of inversion symmetry in SrSi$_{2}$, the SOC is expected to lift the spin degeneracy. Having compared the electronic band structure with and without SOC, we find that the overall nature of band structure is identical due to the involvement of light $Z$ elements of SrSi$_{2}$. The inclusion of  spin-orbit coupling leads to a gapless state at the Fermi level identical to the case when the spin-orbit coupling is not taken into account.

\subsection{Strain influenced electronic structure of SrSi$_{2}$}\label{sec3.2} 
 Initially, we investigated the effect of triaxial strain, i.e., the lattice parameter ``a'' was reduced uniformly in all three directions. We found that with the application of triaxial strain, SrSi$_{2}$ retained its WSM behaviour, albeit the node separation increased progressively as the compressive strain was increased. The above was reported earlier by Bahadur {et al} \cite{ref23}. The authors also report on the observation of a semimetal to trivial insulator transition  with tensile triaxial  strain\cite{ref23}. Hence, SrSi$_{2}$ remains  semimetallic with the application of compressive triaxial strain and becomes a trivial insulator with the application of tensile strain. Next,  we  applied  uniaxial strain on the crystalline $c$-axis while keeping lattice parameters along $a$ and $b$ axis  unchanged [figure~\ref{fig2}(c)]. This geometry enforces anisotropy in an otherwise cubic SrSi$_{2}$, thereby lowering the symmetry of lattice from cubic to tetragonal. We progressively increased the strain (compressive) and, at 5\% value, we notice a slight opening of the band gap of SrSi$_{2}$ along the high-symmetry direction $\Gamma$-X. We conclude that the introduction of anisotropy in the lattice triggers the opening of the band gap near Fermi level which pushes the system into an insulating state.
  To address the question  whether this insulating phase holds at increased values of strain, we applied a compressive strain till 10\% and found that the gapped state is persistent although the band gap increases slightly. The electronic band structures shown in figure~\ref{fig3}(c-f) are corresponding to two selected values of the strain out of several values:  5\% (where the gapped state first appears) and 10\% uniaxial strain,  with lattice parameter $c = 6.236$~{\AA} and $c = 5.907$~{\AA}, respectively. Figure~\ref{fig3} c) and d) shows the bulk band structures for 5\% uniaxial strain and figures~\ref{fig3}e) and f) show the same for 10\% uniaxial strain along the $z$-axis, with and without SOC. The observed band gap was 51  meV and 62 meV for 5\% and 10\% strained structure, respectively. 
  After the inclusion of SOC, the bandgap reduction was observed, from 51 meV to
  36 meV and from 62 meV to 46 meV, for 5\% and 10\% strained structure, respectively.

%
  Further, we computed the p and d orbitals projected band structures for 5\% [figure~\ref{main_label}(a--f)] and 10\% 
  strain [figure ~\ref{main_label}(g--l)], respectively. The red, green and blue coloured circles in figure~\ref{main_label} represent the $p_{x}$, p$_{y}$ and $p_{z}$  orbitals of Si and cyan, magenta and yellow coloured circles represent $d_{xy}$, $d_{yz}$ and $d_{xz}$ orbitals
  of Sr.
 From the figure, we observe that the contribution of  $p_{y}$ and $p_{z}$ orbitals is relatively large near  the Fermi level as compared to  $p_{x}$ orbital. The figure also shows that t$_{2g}$ orbitals (including $d_{xy}$, $d_{yz}$ and $d_{xz}$ orbitals) are present closer to the Fermi level (marked as zero) while e$_{g}$ orbitals ($d_{x^{2}-y^{2}}$ and $d_{z^{2}}$) reside far away from Fermi level. A similar trend was observed in the case of 10\% strain. 
 The total and partial electronic density of states (DOS) of SrSi$_{2}$ for bulk and strained structures are calculated and illustrated in figure~\ref{some-label}(a) and figure~\ref{some-label}(b) and \ref{some-label}(c), respectively, where the Fermi level is set to 0 eV. 
   The contributions of $s$, $p$, $d$ orbitals of Sr and $s$, $p$ orbitals of Si are displayed. For bulk SrSi$_{2}$, a narrow intense peak near the Fermi level is observed but the  same feature was not observed for strained structures. In the vicinity of the Fermi level, the calculated values of partial DOS show that the contributions of  Sr $d$-states are predominant whereas the contributions of Si states are comparatively smaller. The large contributions of $d$-states near the Fermi level are indicative of large  electronic correlations and higher localization than the other known topological materials, which makes SrSi$_{2}$ a good platform for illustrating the effects of correlations on topology.
  We also examined the dynamical stability of strain impacted SrSi$_{2}$ by computing the phonon dispersion as shown in figure~\ref{fig6}a) and b). 
  There are total 36 phonon modes in the vibrational spectra, out of which 3 are the acoustical and rest are the optical modes.
  The absence of any imaginary frequencies along the high-symmetry points indicates that the structure is dynamically stable at the values of strain considered in this work. 
  

%
\subsection{Band inversion and Z$_{2}$ invariants:} \label{sec3.3}

\begin{figure}[!b]
	\centering
	\includegraphics[width=12cm, height=8cm]{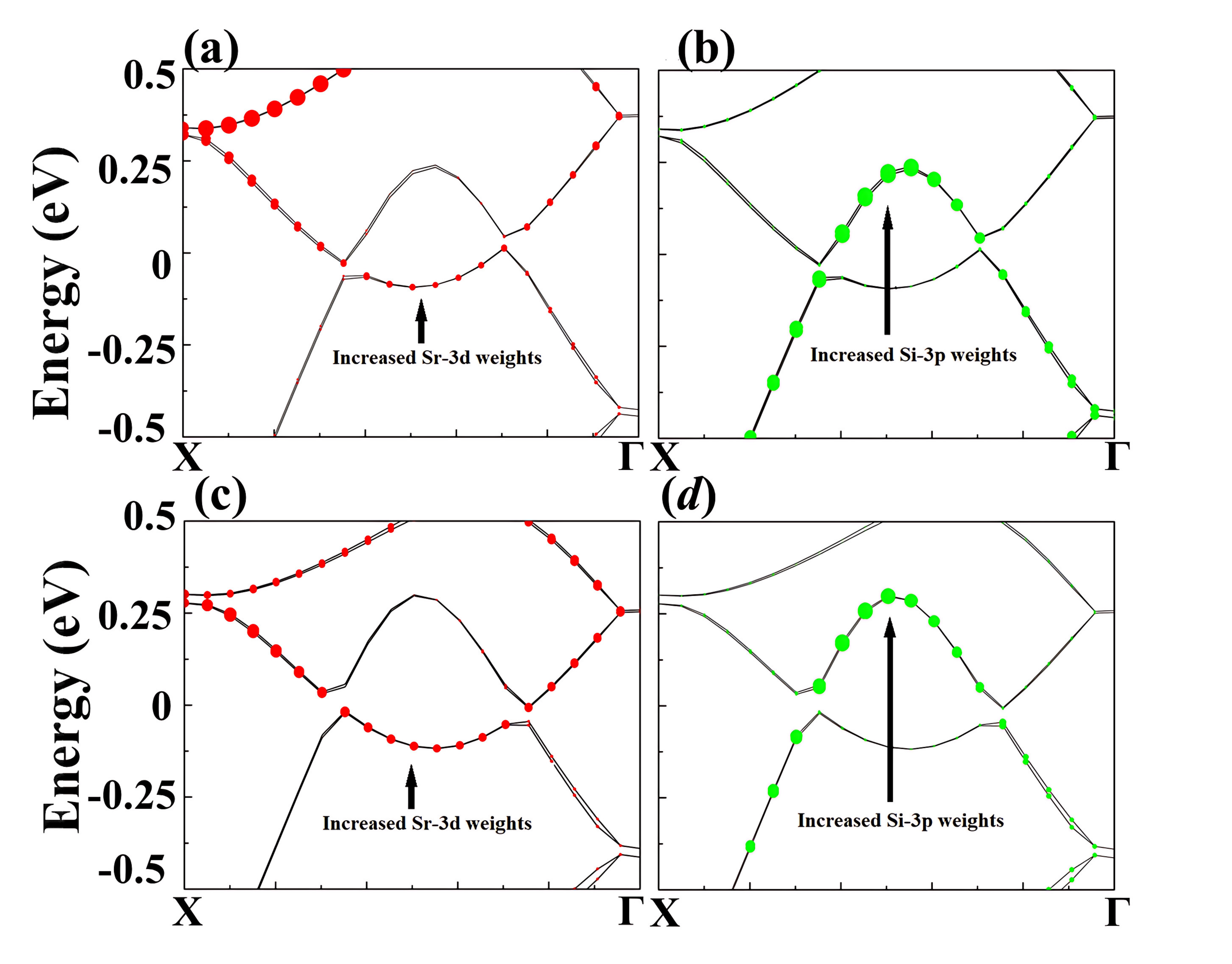}
	\caption{(Colour online) 
		$d-p$ band inversion is seen in (a) Sr-$d$ orbital and (b) Si-$p$ orbital for 5\% strained structure; (c) Sr-$d$ orbital and (d) Si-$p$ orbital for 10\% strained structure.
	}
	\label{fig7}
\end{figure}

 Topologically non-trivial materials could be well identified through the band-inversion phenomenon.  Due to the relativistic impact from heavy elements,  the s-orbits may be pushed underneath the $p$-orbits leading to an inverted band character known as band-inversion. Hence, in order to check whether the insulating phase obtained at 5\% strain is trivial or non-trivial, we study the signatures of band inversion.  In this work, we predict  \textit{d-p} band inversion in strain induced SrSi$_{2}$. The  \textit{d-p} band inversion is unconventional as compared to band inversions found in previous topological insulators. In Bi$_{2}$Se$_{3}$, topological band inversion occurs only due to $p$-orbitals \cite{ref43}. We find that, starting from 5\% strain, the band structure reveals a band inversion along $\Gamma$-X direction when SOC is included in the calculation. The inverted band structure is obtained between the Sr-$d$ orbital and Si-$p$ orbitals as  shown in figure~\ref{fig7} (for 5\% and 10\% strain), which may indicate TI phase. Figure~\ref{fig7}(a) and \ref{fig7}(c) shows the Sr $d$-states for 5\% and 10\% strained SrSi$_{2}$. Here, the radii of red circles correspond to the proportion of Sr-$d$ electrons.We observe that the weight of Sr-$d$ orbital is higher for the valence band compared to the conduction band, in contrast to the ground state (not shown here) where CB is dominated by Sr $d$-states. In figure~\ref{fig7}(b) and \ref{fig7}(d) we  plotted Si $p$-states (the radii of green circles correspond to the proportion of Si-$p$ electrons) for 5\% and 10\% strained SrSi$_2$, respectively, and find that the weight of Si-$p$ orbital is higher in the conduction band as compared to the valence band, while for the ground state, the VB is composed of Si $p$-states. 
%
 In 5\% strain the band inversion happens in Sr $d$-states (specifically in $d_{yz}$ and $d_{xz}$ orbitals as seen in figure~\ref{main_label}) and Si $p$-states (specifically in $p_y$ and $p_z$ orbitals as seen in figure~\ref{main_label}), whereas in 10\% strain, the band inversion happens in Sr $d$-states (specifically in $d_{yz}$ and $d_{xz}$ orbitals) and Si $p$-states (specifically in $p_{x}$ and $p_{y}$ orbitals).
  Thus, the  \textit{d-p} band inversion is apparent in the strained structures. Hence, we conclude that there may be a  topological semimetal to topological insulator transition at 5\% uniaxial strain. In many well-studied topological materials like HgTe/CdTe quantum wells and Bi$_{2}$Se$_{3}$, significant interplay between only $s$ and $p$ orbitals can be observed. The $d-p$ inversion observed in the present work for SrSi$_{2}$, was earlier  observed in a few topological materials such as bismuth-based skutterudites \cite{ref44}.

Band inversion is an essential but not sufficient condition indicating the presence of a topologically non-trivial phase. In order to further confirm the topological insulating phase, it is necessary to evaluate the topological invariants: four Z$_{2}$ indices $\nu_{0}$: ($\nu_{1}$,$\nu_{2}$,$\nu_{3}$), defined by Fu, Kane and Mele \cite{ref45,ref46}. Here, we can obtain Z$_{2}$ topological index  $\nu_{0}$: ($\nu_{1}$,$\nu_{2}$,$\nu_{3}$) from the  calculations of six time reversal invariant planes, i.e.,  (a) k$_{1}=0$; (b) k$_{1}=0.5$; (c) k$_{2}=0$; (d) k$_{2}=0.5$; (e) k$_{3}=0$; (f) k$_{3}=0.5$. The topological invariant, $\nu_{0}$ and $\nu_{i}$, is calculated as $\nu_{0}=(Z_{2}(k_{i}=0)+Z_{2}(k_{i}=0.5))\mod 2$ and $\nu_{i}=Z_{2}(k_{i}=0.5)$ \cite{ref40}. The material is categorized as strong TI if $\nu_{0}$ = 1 and a weak insulator if $\nu_{0}$ = 0 \cite{ref32}. Using the above method the Z$_{2}$ invariants for 5\% and 10\% strained structures were found to be (1, 100) and (1, 111), respectively.  Hence, the topological invariants reveal that SrSi$_{2}$ under uniaxial strain (5\% and 10\%) is a strong topological insulator.

\subsection{Surface states}\label{sec3.4}
\begin{figure}[!h]
	\centering
	\includegraphics[width=11cm, height=10cm]{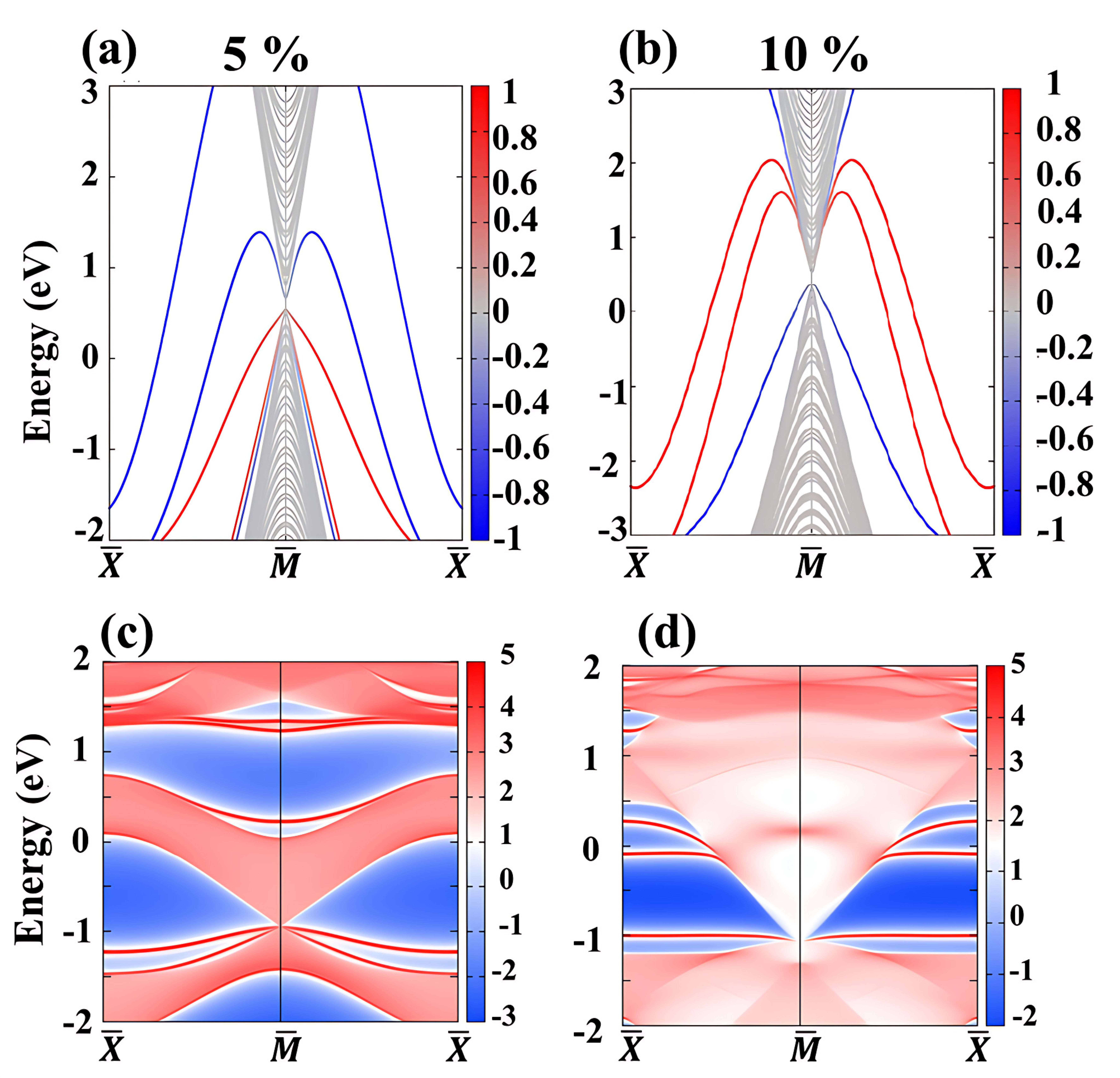}
	\caption{(Colour online) a) and b) Slab band structure; c) and d) surface states of 5\% and 10\% strained structure of SrSi$_{2}$.}
	%
\label{fig8}
\end{figure}

The presence of surface states is one of the most crucial indications of the topological insulating phase. Surface states are the extraordinary states that appear inside the bulk energy gap and permit a metallic conduction on the surface of the topological insulator.  To explore the non-trivial surface states, we performed calculations for the projected band structure of a strained SrSi$_{2}$ on the (001) surface. We  utilized the Wannier Tool \cite{ref47} package based on a tight-binding model with Wannier wave functions built from the plane wave solution. In order to study the surface states, we consider the slab structure which is made up of a slab of 25 layers of SrSi$_{2}$ along (001) surface.  The slab band structure for strain-induced SrSi$_{2}$ is shown in figure~{\ref{fig8}}(a) and (b) for 5\% and 10\% uniaxial strain. The gapped bulk states are clearly seen in the slab band structure. This again confirms the prediction about \textit{d-p} band inversion yielding a topological insulating phase in strained  SrSi$_{2}$. Figure~{\ref{fig8}}(c) and (d) illustrates the topologically protected metallic surface states along $\bar{X}-\bar{M}-\bar{X}$ direction over the surface Brillouin  zone, near Fermi level.

\section{Conclusion}\label{sec4}
 In conclusion, using density functional theory-based calculations, we  have demonstrated the suitability of external uniaxial strain  to control the band structure and topological properties of the semimetallic SrSi$_{2}$.   We examined the modifications in the band structure of SrSi$_{2}$ by progressively increasing the uniaxial compressive strain along the crystalline $c$-axis ($z$-direction). At 5\% and beyond, a slight opening of the bandgap is observed. To confirm whether this insulating state is trivial or non-trivial, we calculated topological invariants and the surface band structure. We observe  topologically protected surface states on the (001) surface of strained SrSi$_{2}$. Hence, we conclude that strain-induced SrSi$_{2}$ is a strong topological insulator and undergoes a TSM $\leftrightarrow$ TI phase transition under the effect of uniaxial strain. Our findings reveal that the topological behaviour of SrSi$_{2}$ could be well controlled with uniaxial strain and may prospectively trigger future interests in developing SrSi$_{2}$-based electronic devices.

\subsection{Acknowledgements}
Author PS would like to thank NPSF C-DAC Pune for providing HPC facility.



\begin{thebibliography}{99}
	
	\bibitem{ref1}
	Qi~X.-L., Zhang~S.-C., Phys. Today, 2010, \textbf{63}, No.~1, 33--38,
	\doi{10.1063/1.3293411}.
	
	\bibitem{ref2}
	Hasan~M.~Z., Kane~C.~L., Rev. Mod. Phys., 2010, \textbf{82}, 3045--3067,
	\doi{10.1103/RevModPhys.82.3045}.
	
	\bibitem{ref3}
	Moore~J.~E., Nature, 2010, \textbf{464}, No. 7286, 194--198,
	\doi{10.1038/nature08916}.
	
	\bibitem{ref4}
	Young~S.~M., Zaheer~S., Teo~J. C.~Y., Kane~C.~L., Mele~E.~J., Rappe~A.~M.,
	Phys. Rev. Lett., 2012, \textbf{108}, 140405,
	\doi{10.1103/PhysRevLett.108.140405}.
	
	\bibitem{ref5}
	Wang~Z., Sun~Y., Chen~X.-Q., Franchini~C., Xu~G., Weng~H., Dai~X., Fang~Z.,
	Phys. Rev. B, 2012, \textbf{85}, 195320, \doi{10.1103/PhysRevB.85.195320}.
	
	\bibitem{ref6}
	Wan~X., Turner~A.~M., Vishwanath~A., Savrasov~S.~Y., Phys. Rev. B, 2011,
	\textbf{83}, 205101, \doi{10.1103/PhysRevB.83.205101}.
	
	\bibitem{ref7}
	Gao~H., Venderbos~J.~W., Kim~Y., Rappe~A.~M., Annu. Rev. Mater. Res., 2019, \textbf{49}, No.~1, 153--183,
	\doi{10.1146/annurev-matsci-070218-010049}.
	
	\bibitem{ref8}
	Bernevig~B.~A., Hughes~T.~L., Zhang~S.-C., Science, 2006, \textbf{314}, No.
	5806, 1757--1761, \\\doi{10.1126/science.1133734}.
	
	\bibitem{ref9}
	K{\"o}nig~M., Wiedmann~S., Br{\"u}ne~C., Roth~A., Buhmann~H., Molenkamp~L.~W.,
	Qi~X.-L., Zhang~S.-C., Science, 2007, \textbf{318}, 766--770,
	\doi{10.1126/science.1148047}.
	
	\bibitem{ref10}
	Zhang~H., Liu~C.-X., Qi~X.-L., Dai~X., Fang~Z., Zhang~S.-C., Nat. Phys.,
	2009, \textbf{5}, 438--442, \doi{10.1038/nphys1270}.
	
	\bibitem{ref11}
	Hsieh~D., Xia~Y., Qian~D., Wray~L., Meier~F., Dil~J.~H., Osterwalder~J.,
	Patthey~L., Fedorov~A.~V., Lin~H., et~al., Phys. Rev. Lett., 2009, \textbf{103}, 146401,
	\doi{10.1103/PhysRevLett.103.146401}.
	
	\bibitem{ref12}
	Jin~H., Rhim~S.~H., Im~J., Freeman~A.~J., Sci. Rep., 2013, \textbf{3},
	1651, \doi{10.1038/srep01651}.
	
	\bibitem{ref13}
	Lin~H., Wray~L.~A., Xia~Y., Xu~S., Jia~S., Cava~R.~J., Bansil~A., Hasan~M.~Z.,
	Nat. Mater., 2010, \textbf{9}, 546--549.
	
	\bibitem{ref14}
	Chadov~S., Qi~X., K{\"u}bler~J., Fecher~G.~H., Felser~C., Zhang~S.~C., Nat. Mater., 2010, \textbf{9}, No.~7, 541--545, \doi{10.1038/nmat2770}.
	
	\bibitem{ref15}
	Franz~M., Nat. Mater., 2010, \textbf{9}, No.~7, 536--537,
	\doi{10.1038/nmat2783}.
	
	\bibitem{ref16}
	Yan~B., Liu~C.-X., Zhang~H.-J., Yam~C.-Y., Qi~X.-L., Frauenheim~T.,
	Zhang~S.-C., Europhys. Lett., 2010, \textbf{90}, 37002, \doi{10.1209/0295-5075/90/37002}.
	
	\bibitem{ref17}
	Wang~Z., Weng~H., Wu~Q., Dai~X., Fang~Z., Phys. Rev. B, 2013, \textbf{88},
	125427, \doi{10.1103/PhysRevB.88.125427},
\url{https://link.aps.org/doi/10.1103/PhysRevB.88.125427}.
	
	\bibitem{ref18}
	Xu~S.-Y., Alidoust~N., Belopolski~I., Yuan~Z., Bian~G., Chang~T.-R., Zheng~H.,
	Strocov~V.~N., Sanchez~D.~S., Chang~G., et~al., Nat. Phys., 2015,
	\textbf{11}, 748--754, \doi{10.1038/nphys3437}.
	
	\bibitem{ref19}
	Xu~S.-Y., Belopolski~I., Alidoust~N., Neupane~M., Bian~G., Zhang~C., Sankar~R.,
	Chang~G., Yuan~Z., Lee~C.-C., et~al., Science, 2015, \textbf{349}, 613--617,
	\doi{10.1126/science.aaa9297}.
	
	\bibitem{ref20}
	Lv~B.~Q., Weng~H.~M., Fu~B.~B., Wang~X.~P., Miao~H., Ma~J., Richard~P.,
	Huang~X.~C., Zhao~L.~X., Chen~G.~F., et~al., Phys.
	Rev. X, 2015, \textbf{5}, 031013, \doi{10.1103/PhysRevX.5.031013}.
	
	\bibitem{ref21}
	Lv~B.~Q., Xu~N., Weng~H.~M., Ma~J.~Z., Richard~P., Huang~X.~C., Zhao~L.~X.,
	Chen~G.~F., Matt~C.~E., Bisti~F., et~al., Nat. Phys., 2015, \textbf{11}, 724--727,
	\doi{10.1038/nphys3426}.
	
	\bibitem{ref22}
	Yang~L.~X., Liu~Z.~K., Sun~Y., Peng~H., Yang~H.~F., Zhang~T., Zhou~B.,
	Zhang~Y., Guo~Y.~F., Rahn~M., et~al., Nat. Phys., 2015, \textbf{11}, 728--732,
	\doi{10.1038/nphys3425}.
	
	\bibitem{ref23}
	Singh~B., Chang~G., Chang~T.-R., Huang~S.-M., Su~C., Lin~M.-C., Lin~H.,
	Bansil~A., Sci. Rep., 2018, \textbf{8}, 10540,
	\doi{10.1038/s41598-018-28644-y}.
	
	\bibitem{ref24}
	Huang~S.-M., Xu~S.-Y., Belopolski~I., Lee~C.-C., Chang~G., Chang~T.-R.,
	Wang~B., Alidoust~N., Bian~G., Neupane~M., et~al., Proc. Natl. Acad. Sci. U.S.A., 2016, \textbf{113}, 1180--1185,
	\doi{10.1073/pnas.1514581113}.
	
	\bibitem{ref25}
	Wang~Z., Vergniory~M.~G., Kushwaha~S., Hirschberger~M., Chulkov~E.~V.,
	Ernst~A., Ong~N.~P., Cava~R.~J., Bernevig~B.~A., Phys. Rev. Lett., 2016,
	\textbf{117}, 236401, \doi{10.1103/PhysRevLett.117.236401}.
	
	\bibitem{ref26}
	Chang~G., Xu~S.-Y., Zheng~H., Singh~B., Hsu~C.-H., Bian~G., Alidoust~N.,
	Belopolski~I., Sanchez~D.~S., Zhang~S., Lin~H., Hasan~M.~Z., Sci. Rep., 2016, \textbf{6}, 38839, \doi{10.1038/srep38839}.
	
	\bibitem{ref27}
	Lv~B.~Q., Feng~Z.-L., Xu~Q.-N., Gao~X., Ma~J.-Z., Kong~L.-Y., Richard~P.,
	Huang~Y.-B., Strocov~V.~N., Fang~C., et~al.,
	Nature, 2017, \textbf{546}, 627--631, \doi{10.1038/nature22390}.
	
	\bibitem{ref28}
	Ma~J.-Z., He~J.-B., Xu~Y.-F., Lv~B.~Q., Chen~D., Zhu~W.-L., Zhang~S.,
	Kong~L.-Y., Gao~X., Rong~L.-Y., et~al., Nat. Phys., 2018, \textbf{14},
	349--354, \doi{10.1038/s41567-017-0021-8}.
	
	\bibitem{ref29}
	Neupane~M., Belopolski~I., Hosen~M.~M., Sanchez~D.~S., Sankar~R., Szlawska~M.,
	Xu~S.-Y., Dimitri~K., Dhakal~N., Maldonado~P., et~al., Phys. Rev. B, 2016,
	\textbf{93}, 201104, \doi{10.1103/PhysRevB.93.201104}.
	
	\bibitem{ref30}
	Shao~D., Ruan~J., Wu~J., Chen~T., Guo~Z., Zhang~H., Sun~J., Sheng~L., Xing~D.,
	Phys. Rev. B, 2017, \textbf{96}, 075112, \doi{10.1103/PhysRevB.96.075112}.
	
	\bibitem{ref31}
	Juneja~R., Shinde~R., Singh~A.~K., J. Phys. Chem. Lett.,
	2018, \textbf{9}, 2202--2207, \doi{10.1021/acs.jpclett.8b00646}.
	
	\bibitem{PhysRevB.91.035311}
	K\"ufner~S., Bechstedt~F., Phys. Rev. B, 2015, \textbf{91}, 035311,
	\doi{10.1103/PhysRevB.91.035311}.
	
	\bibitem{PhysRevLett.106.126803}
	Br\"une~C., Liu~C.~X., Novik~E.~G., Hankiewicz~E.~M., Buhmann~H., Chen~Y.~L.,
	Qi~X.~L., Shen~Z.~X., Zhang~S.~C., Molenkamp~L.~W., Phys. Rev. Lett., 2011,
	\textbf{106}, 126803, \doi{10.1103/PhysRevLett.106.126803}.
	
	\bibitem{PhysRevB.85.195114}
	Feng~W., Zhu~W., Weitering~H.~H., Stocks~G.~M., Yao~Y., Xiao~D., Phys. Rev. B,
	2012, \textbf{85}, 195114, \doi{10.1103/PhysRevB.85.195114}.
	
	\bibitem{Sklyadneva2016}
	Sklyadneva~I.~Y., Rusinov~I.~P., Heid~R., Bohnen~K.-P., Echenique~P.~M.,
	Chulkov~E.~V., Sci. Rep., 2016, \textbf{6}, 24137,
	\doi{10.1038/srep24137}.
	
	\bibitem{PhysRevLett.117.146402}
	Zhou~Y., Lu~P., Du~Y., Zhu~X., Zhang~G., Zhang~R., Shao~D., Chen~X., Wang~X.,
	Tian~M., et~al., Phys. Rev.
	Lett., 2016, \textbf{117}, 146402, \doi{10.1103/PhysRevLett.117.146402}.
	
	\bibitem{ref35}
	Zeng~Q., Sun~H., Shen~J., Yao~Q., Zhang~Q., Li~N., Jiao~L., Wei~H., Felser~C.,
	Wang~Y., et~al., Adv. Quantum Technol., 2022, \textbf{5},
	2100149, \doi{10.1002/qute.202100149}.
	
	\bibitem{ref32}
	Winterfeld~L., Agapito~L.~A., Li~J., Kioussis~N., Blaha~P., Chen~Y.~P., Phys.
	Rev. B, 2013, \textbf{87}, 075143, \doi{10.1103/PhysRevB.87.075143}.
	
	\bibitem{ref33}
	Mutch~J., Chen~W.-C., Went~P., Qian~T., Wilson~I.~Z., Andreev~A., Chen~C.-C.,
	Chu~J.-H., Sci. Adv., 2019, \textbf{5}, \doi{10.1126/sciadv.aav9771}.
	
	\bibitem{ref34}
	Huang~Z.-Q., Hsu~C.-H., Chuang~F.-C., Liu~Y.-T., Lin~H., Su~W.-S., Ozolins~V.,
	Bansil~A., New J. Phys., 2014, \textbf{16}, 105018,
	\doi{10.1088/1367-2630/16/10/105018}.
	
	\bibitem{PhysRevB.102.035132}
	Schindler~C., Noky~J., Schmidt~M., Felser~C., Wosnitza~J., Gooth~J., Phys. Rev.
	B, 2020, \textbf{102}, 035132, \doi{10.1103/PhysRevB.102.035132}.
	
	\bibitem{ref36}
	Giannozzi~P., Baroni~S., Bonini~N., Calandra~M., Car~R., Cavazzoni~C.,
	Ceresoli~D., Chiarotti~G.~L., Cococcioni~M., Dabo~I.,   et~al., J. Phys.: Condens. Matter, 2009, \textbf{21}, No.~39, 395502,
	\doi{10.1088/0953-8984/21/39/395502}.
	
	\bibitem{ref37}
	Pizzi~G., Vitale~V., Arita~R., Bl{\"u}gel~S., Freimuth~F., G{\'e}ranton~G.,
	Gibertini~M., Gresch~D., Johnson~C., Koretsune~T.,  et~al., J. Phys.: Condens. Matter, 2020, \textbf{32}, No.~16, 165902,\\
	\doi{10.1088/1361-648X/ab51ff}.
	
	\bibitem{ref38}
	Thygesen~K.~S., Hansen~L.~B., Jacobsen~K.~W., Phys. Rev. Lett., 2005,
	\textbf{94}, 026405, \doi{10.1103/PhysRevLett.94.026405}.
	
	\bibitem{ref39}
	Souza~I., Marzari~N., Vanderbilt~D., Phys. Rev. B, 2001, \textbf{65}, 035109,
	\doi{10.1103/PhysRevB.65.035109}.
	
	\bibitem{ref40}
	Mostofi~A.~A., Yates~J.~R., Lee~Y.-S., Souza~I., Vanderbilt~D., Marzari~N.,
	Comput. Phys. Commun., 2008, \textbf{178}, No.~9, 685--699,
	\doi{10.1016/j.cpc.2007.11.016}.
	
	\bibitem{ref41}
	Togo~A., Tanaka~I., Scr. Mater., 2015, \textbf{108}, 1--5,
	\doi{10.1016/j.scriptamat.2015.07.021}.
	
	\bibitem{ref42}
	Hashimoto~K., Kurosaki~K., Imamura~Y., Muta~H., Yamanaka~S., J. Appl. Phys., 2007, \textbf{102}, 063703, \doi{10.1063/1.2778747}.
	
		\bibitem{ref42ol}
	Evers~J., J. Solid State Chem., 1978, \textbf{24}, No.~2, 199-–207, \doi{10.1016/0022-4596(78)90010-5}.
	
	\bibitem{ref43}
	Chege~S., Ningi~P., Sifuna~J., Amolo~G.~O., AIP Adv., 2020, \textbf{10},
	095018, \doi{10.1063/5.0022525}.
	
	\bibitem{ref44}
	Yang~M., Liu~W.~M., Sci. Rep., 2014, \textbf{4}, 5131,
	\doi{10.1038/srep05131}.
	
	\bibitem{ref45}
	Fu~L., Kane~C.~L., Phys. Rev. B, 2007, \textbf{76}, 045302,
	\doi{10.1103/PhysRevB.76.045302}.
	
	\bibitem{ref46}
	Fu~L., Kane~C.~L., Mele~E.~J., Phys. Rev. Lett., 2007, \textbf{98}, 106803,
	\doi{10.1103/PhysRevLett.98.106803}.
	
	\bibitem{ref47}
	Wu~Q., Zhang~S., Song~H.-F., Troyer~M., Soluyanov~A.~A., Comput. Phys. Commun., 2018, \textbf{224}, 405--416,
	\doi{10.1016/j.cpc.2017.09.033}.
	
\end{thebibliography}

\newpage
\ukrainianpart

\title[Топологічні фазові переходи в SrSi$_{2}$]%
{Керовані тиском топологічні фазові переходи зі стану ізолятора Вейля у напівметал SrSi$_{2}$}
\author[А. Шенде, С. Кумар Гупта, А. Коре, П. Сінгх]{А. Шенде, С. Кумар Гупта, А. Коре, П. Сінгх}
\address{Фізичний факультет, Національний інститут технології Вісвесварая, С. Амбасарі Роуд, Нагпур, 440010, Махараштра, Індія}

\makeukrtitle

\begin{abstract}
	З використанням першопринципних розрахунків на основі методу функціонала густини, показано перелаштування електронної структури напівметалу Вейля SrSi$_{2}$ під дією зовнішньої одновісної деформації. Одновісна деформація сприяє відкриттю забороненої зони вздовж напрямку $\Gamma$-X та подальшій інверсії зони між орбіталями Si p та Sr d. Z$_{2}$-інваріанти та поверхневі стани переконливо показують, що при одновісній деформації SrSi$_{2}$ є хорошим топологічним ізолятором. Таким чином, одновісна деформація переводить напівметалічний SrSi$_{2}$ у стан топологічного ізолятора з відповідною забороненою зоною, що відповідає фазовому переходу ``напівметал-топологічний ізолятор''. Отримані результати підтверджують придатність одновісної деформації для контролю над топологічними фазовими переходами та топо\-ло\-гіч\-ни\-ми станами в SrSi$_{2}$.
	\keywords топологічні напівметали, топологічні ізолятори, теорія функціоналу густини, електронна структура, фазові переходи
\end{abstract}
\end{document}